\documentclass[twocolumn]{aastex62}
\usepackage{amsmath,amstext}
\usepackage{amsmath, amsthm, amssymb, amsfonts}
\usepackage{graphicx}
\usepackage{rotating}
\definecolor{myred}{cmyk}{0, 0.7808, 0.4429, 0.1412}

\shorttitle{Detecting Vortices in Protoplanetary Disks}
\shortauthors{Huang P. et al.}

\begin{document}

\title{Identifying Anticyclonic Vortex Features
Produced by the Rossby Wave Instability
in Protoplanetary Disks}

\correspondingauthor{Pinghui Huang}
\email{phhuang@pmo.ac.cn}

\author{Pinghui Huang}
\affiliation{CAS Key Laboratory of Planetary Sciences, Purple Mountain Observatory, Chinese Academy of Sciences,
Nanjing 210008, China}
\affiliation{University of Chinese Academy of Sciences, Beijing 100049, China}
\affiliation{Theoretical Division, Los Alamos National Laboratory, Los Alamos, NM 87545, USA}
\author{Andrea Isella}
\affiliation{Department of Physics \& Astronomy, Rice University, 6100 Main Street, Houston, TX 77005, USA}
\author{Hui Li}
\affiliation{Theoretical Division, Los Alamos National Laboratory, Los Alamos, NM 87545, USA}
\author{Shengtai Li}
\affiliation{Theoretical Division, Los Alamos National Laboratory, Los Alamos, NM 87545, USA}
\author{Jianghui Ji}
\affiliation{CAS Key Laboratory of Planetary Sciences, Purple Mountain Observatory, Chinese Academy of Sciences,
Nanjing 210008, China}

\begin{abstract}
Several nearby protoplanetary disks have been observed to display large scale crescents in the (sub)millimeter 
dust continuum emission. One interpretation is that these structures correspond to anticyclonic 
vortices generated by the Rossby wave instability within the gaseous disk. Such vortices have local gas 
over-densities and are expected to concentrate dust particles with Stokes number around unity. This process might 
catalyze the formation of planetesimals. Whereas recent observations 
showed that dust crescent are indeed regions where millimeter-size particles have abnormally high concentration 
relative to the gas and smaller grains, no observations have yet shown that the gas within the crescent region 
counter-rotates with respect to the protoplanetary disk. Here we investigate the detectability of 
anticyclonic features through 
measurement of the line-of-sight component of the gas velocity obtained with ALMA. We carry out 2D 
hydrodynamic simulations and 3D radiative transfer calculation of a protoplanetary disk characterized by a vortex 
created by the tidal interaction with a massive planet. As a case study, the disk parameters are chosen to mimic the 
IRS~48 system, which has the most prominent crescent observed to date. We generate synthetic ALMA observations of 
both the dust continuum and $^{12}$CO emission around the frequency of 345 GHz. We find that the anticyclonic features of vortex are weak but can be detected if both the source and the observational setup are properly 
chosen. We provide a recipe for maximizing the probability to detect such vortex features and present 
an analysis procedure to infer their kinematic properties.

\end{abstract}

\section{Introduction}
Protoplanetary systems such as IRS48 \citep{van2013major}, LkH$\alpha$~330 \citep{isella2013azimuthal}, HD~142527 \citep{muto2015significant}, MWC~758 \citep{2010ApJ...725.1735I},  and SAO~206462 \citep{2014ApJ...783L..13P} exhibit large cavities and prominent crescents in the (sub-)millimeter dust continuum emission. In systems where spatially resolved multi-wavelength observations of both dust and gas emission exist, namely IRS~48, HD~142527, and MWC 758, the continuum crescents appear to originate from the azimuthal and radial concentration of solid particles toward local maxima of the gas pressure \citep{2015ApJ...812..126C,boehler2017close,2015ApJ...810L...7V}. Such concentration is believed to result from the decoupling between dust and gas, and subsequent migration of solid particles in the direction of the gas pressure gradient \citep[see, e.g.,][]{weidenschilling1977aerodynamics,2013A&A...550L...8B}. This interpretation is supported by the fact that gas-dust decoupling is most efficient for grains with Stokes number around unity, which, for typical densities of protoplanetary disks, correspond indeed to grains that emit mostly at (sub-)millimeter wavelengths.

Whereas the motion of dust particles in gaseous disks is relatively well understood, the origin of gas pressure 
maxima is not. 
A commonly accepted explanation for the existence of asymmetric pressure maxima in protoplanetary disks is that they originate from large scale vortices resulting from hydrodynamic instabilities. 
Specifically, vortices can be generated by  planets \citep{li2005potential,zhu2014particle,fu2014effects} or viscosity transitions \citep{regaly2013trapping,miranda2017long}. 
A massive planet or a sharp radial viscosity transition create a steep radial gradient in the gas density which triggers the Rossby Wave Instability \citep[RWI,][]{lovelace1999rossby,li2000rossby} which at nonlinear stage 
can produce large-scale vortices \citep{li2001rossby,li2005potential,ou2007disk}. 
Numerical simulations show that the formation and evolution of a vortex strongly depend on the disk viscosity. In particular, a vortex can form and survive for hundreds to thousands orbits 
only if the viscosity is low ($\alpha \simeq 10^{-3} \thicksim 10^{-5}$, \citealt{fu2014long}).
Taken at face value, this result is in conflict with the typical $\alpha \simeq 0.01$ required to explain mass accretion rates found in protoplanetary disks \citep{hartmann1998accretion}. However, theoretical models suggest that disk viscosity might 
vary with the distance from the star, and that the innermost disk  regions ($<$1 AU), those responsible for the accretion onto the star, might be more viscous than the disk regions probed by current spatially resolved observations ($>$10~AU). Lacking  direct measurements \citep[but see][]{2017ApJ...843..150F}, the level of viscosity in protoplanetary disks remains highly debated.

In addition to informing about disk viscosity, vortices might be pivotal in the formation of planets: 
they can speed up the dust coagulation and the formation of planetesimals \citep{barge1995did}, 
stop the radial inward drift of dust particles \citep{weidenschilling1977aerodynamics}, and help overcome the  fragmentation barrier \citep{brauer2008coagulation}. Vortices generated at the edges of low ionization regions could perhaps lead to the formation of the planetesimals needed by {\it core accretion} model \citep{pollack1996formation} to form giant planets. Moreover, vortices generated by giant planets could 
perhaps control the formation and migration of rocky planets in their region of influence 
\citep[see, e.g.,][]{liu2018ngvla}.  Studying the presence of vortices in protoplanetary disks is therefore important to understand the architecture of the Solar systems and of the multitude of planetary systems discovered so far.

In this paper, we study the direct detectability of vortices from spatially and spectrally resolved observations of the molecular 
line emission at millimeter-wavelengths. Observations of dust continuum crescents support the vortex hypothesis 
and, when combined with molecular line data, provide indication of the efficiency of the dust concentration process. 
However, direct measurements of the gas vorticity are necessary to i) confirm that the observed dust crescents correspond to gas vortices and ii) study 
the mechanisms related to the vortex formation and evolution. Here, we focus on the scenario in which 
a vortex is created by 
a massive planet and simulate its density and velocity structure through high resolution 2D hydrodynamic calculations 
of the disk-planet interaction. We then perform radiative transfer calculation to generate synthetic images of the dust and CO emission, 
and process them through the ALMA simulator to properly account for observational noise.  Synthetic observations 
are then analyzed to search for the signature of the vortex in the intensity moments of the gas intensity. We find that 
kinematic signatures of a vortex in a disk like IRS~48  can be detected by ALMA. We provide the observational set up 
required to achieve such a detection. Our analysis is complementary and extends the recent study of the effect of planet-disk interaction on the CO line emission presented by \cite{perez2018observability}. Note that our analysis should be applicable to vortices excited via different mechanisms 
as described above, because the typical velocity variations associated with vortices are similar, with their magnitudes 
reaching a fraction of the local sound speed
\citep{li2001rossby}.

The outline of this paper is as follows. In Section 2, we describe our numerical model setup and methods. 
In Section 3, we present the results of the numerical simulations and the synthetic images obtained. 
In section 4, we discuss and summarize our results.

\section{Modeling procedure}

\subsection{Planet-disk interaction model}
\label{sec:vortex_model}

We use the two-fluid hydrodynamic 2D code LA-COMPASS \citep{li2005potential,li2008type,fu2014effects} 
to simulate the formation and evolution of a large scale vortex in a protoplanetary disk caused by the interaction with a massive planet. 
The hydrodynamic model is 
initialized using properties similar to that of the IRS 48 system, though our results apply to protoplanetary disks in general. We adopt a stellar mass $M_\star = 2~\mathrm{M}_\sun$ \citep{van2013major} and  a stellar effective temperature $T_\star = 9400$~K \citep{follette2015seeds}. The initial gas surface density profile is $\Sigma_{gas}(r) =\Sigma_{gas,0}(r/r_0)^{-1}$ where $\Sigma_{gas,0}=1.73$ g cm$^{-2}$ and $r_0 = 35$ au. 
The inner and outer boundaries of the disk are $0.4r_0 = 14$ au and $6.68r_0 = 233.8$ au, respectively. 
Modeling by Bruderer et al. (2014) have indicated that the inner gaseous disk of IRS 48 is depleted between $0.4 - 20$ au. Here we have chosen the initial inner gas disk boundary at $14$ au and, with disk-planet interaction, the inner gas disk will be significantly depleted over time. Overall, we find that our results on vortices do not depend on the choice of the inner disk boundary.
At the beginning of the simulation, gas and dust are well coupled with a gas-to-dust ratio of 100. Dust grains are modeled as compact spheres with a diameter of $s_\mathrm{d} = 0.20\;\mathrm{mm}$ and internal density of $\rho_\mathrm{d} = 1.2\;\mathrm{g\;cm^{-3}}$.
The corresponding Stokes number at $r_0$ is 0.022. The choice of the grain size has little effect on our results and is motivated by the fact that these are the grains that emit most the thermal radiation observed at $\lambda = 2\pi s_d \approx 1$~mm. In particular, we find that our specific choice of $0.2$ mm grain size does not affect the gas density and kinematics of the vortex.

Following the results of \cite{van2013major}, we introduce a 10~$\mathrm{M_J}$ mass planet at $r_0=35$ au from the star, so that $\mu = M_{\mathrm{p}}/M_*=5\times 10^{-3}$ and the planet orbital period equals to about 146 years. We perform hydrodynamic simulations by assuming a locally isothermal equation of state, where the radial profile of the temperature is $T(r) = T_0(r/r_0)^{-1/2}$ and the corresponding radial profile of the sound speed is $c_s=c_{s,0}(r/r_0)^{-1/4}$, with $c_{s,0}=\Omega_0 h_0$, $h_0 = 0.085 r_0$, and $\Omega_0 = \sqrt{GM_\star/r_0^3}$. 
This assumed disk temperature is consistent with that derived from the radiative transfer calculation discussed in the next section. We adopt a 2D uniform polar grid with 3072 cells in both the radial and azimuthal direction. Each cell has therefore a radial size of about 0.07~au, or $h_0/30$.
We adopt a Shakura-Sunyaev viscosity $\alpha$  parameter of $7\times 10^{-5}$ \citep{shakura1973black}, which is the optimal value for generating vortices \citep{fu2014long}. In this paper, we do not consider the feedback of dust on the vortex structure \citep{fu2014effects} nor the planet migration. The viscous heating is also neglected in our simulation because the viscous dissipation rate $D(R)=\frac{9}{8}\nu\Sigma\frac{GM}{R^3}$ \citep{frank2002accretion,chiang2010forming}, $\Sigma$ is gas surface density of disk. If the disk radiates as a blackbody $\sigma T^4 = D(R)$, the fluctuation of temperature caused by the viscous heating is small so that we can ignore the viscous heating even at the vortex region.

\begin{figure*}[!t]
\includegraphics[width=\textwidth]{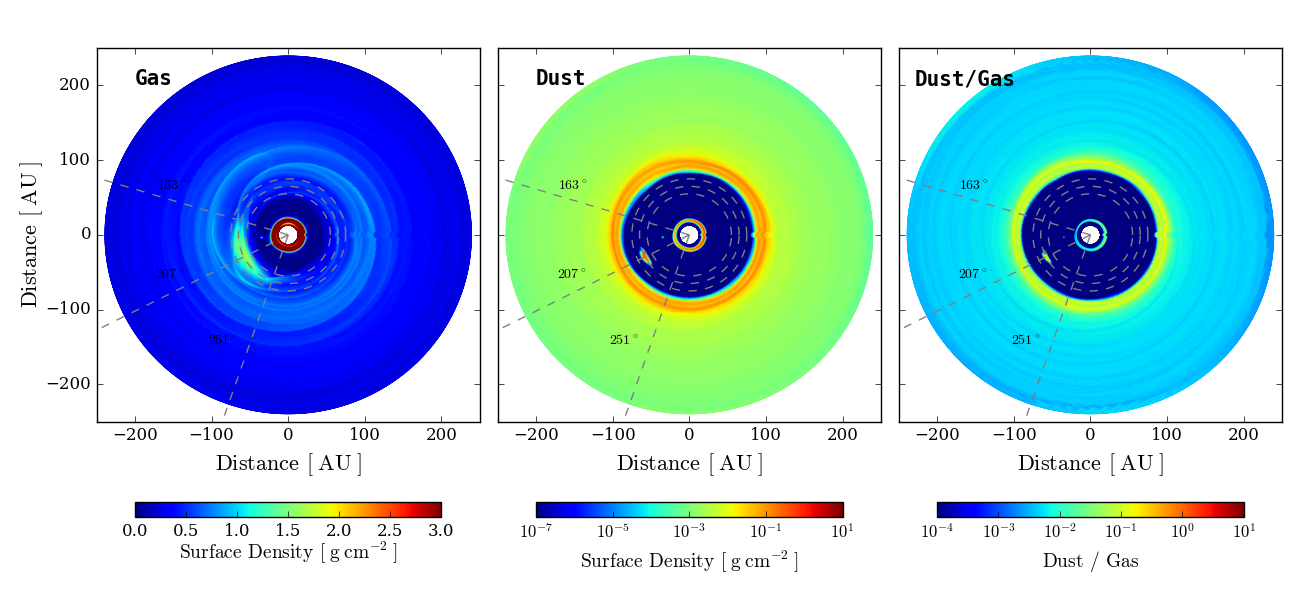}
\caption{Snapshot of the surface density of gas (left) and dust (center) of a disk perturbed by a 10 M$_\mathrm{J}$ planet orbiting at 35 au from a 2 M$_\odot$ star at $10^5$ yr. The planet-disk interaction leads to the formation of a vortex centered at 65 au from the central star at a position angle of 207\arcdeg. The vortex extends azimuthally from $163^\circ$ to $251^\circ$, and radially from 55 au to 75 au. The vortex region in the density profile is marked by the dashed lines and the dashed circles, which indicate radii of  55 au, 65 au, and 75 au, respectively. The right panel shows the dust-to-gas surface density ratio. Planet-disk interaction leads to an enhancement of dust density both within the vortex and at the outer edge of the gap created by the planet. \label{fig:figure1}}

\includegraphics[width=\linewidth]{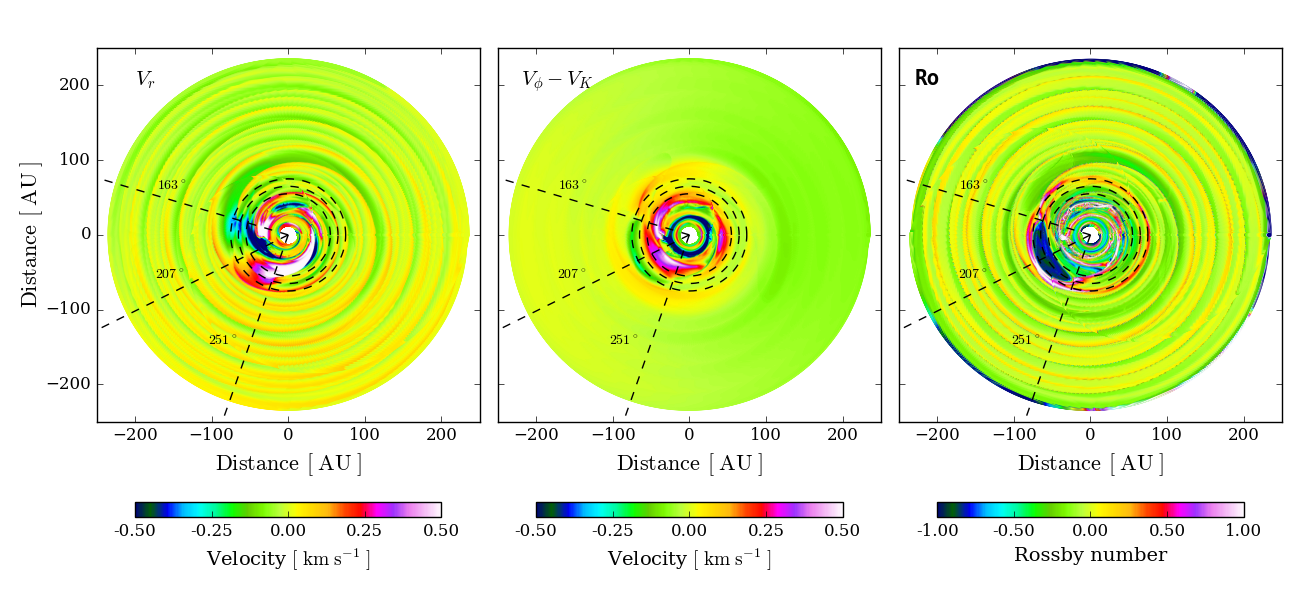}
\caption{Radial (left) and azimuthal (center) components of the gas velocity. The vortex appears as an anticyclone (clockwise rotation) and as a region characterized by negative values of the Rossby number (right). In the central panel, the Keplerian velocity has been subtracted from the gas velocity to better display the perturbations caused by the planet. For reference, the Keplerian velocity at 35 au is 7.12 $\mathrm{km\;s^{-1}}$. \label{fig:figure2}}
\end{figure*}

Figure~\ref{fig:figure1} shows the surface density profile of gas and dust in our two-fluid hydrodynamic simulations after 720 orbits of the planet, or about $10^5$ years. 
The planet carves a deep circular gap in the gas distribution from about 25 au to 45 au (left panel). At the outer edge of the gap, the gas surface density sharply increases by more than a factor  10 within 5 au. This sharp increase causes the formation of a vortex that extends radially from 55 au to 75 au, and azimuthally from 163$^\circ$ to 251$^\circ$ (0$^\circ$ is along the x axis direction). The center of vortex is located at about 65 au from the central star and at an azimthal angle of 207$^\circ$. 
 
The dust distribution is strongly affected by the planet as well. It features a circular gap that is much wider than that is in the gas and extends out to a radius of about 85 au, i.e., well beyond the location of the vortex. Within this gap, our model predicts a  strong accumulation of dust caused by the radial and azimuthal trapping of solid particles toward the center of the vortex. As a result of the disk-planet interaction and of dust migration toward gas pressure maxima, the dust-to-gas ratio increases by about a factor of ten at the outer edge on the dust gap, and by almost a factor of 100 within the vortex. Whereas at the center of the vortex the model predicts a dust-to-gas ratio close to unity, it is important to note that this value is likely overestimated due to the lack of dust feedback which might quench the concentration of solid particles \citep{fu2014effects}.

Figure~\ref{fig:figure2} shows the gas velocity $\vec{v}_g$ along the $r$ and $\phi$ directions relative to the Keplerian motion $\vec{v}_K$, as well as the Rossby number of the gas, $\mathrm{Ro}=2[\nabla \times (\vec{v}_g - \vec{v}_K)]_z /\Omega_K$. The main perturbation induced by the planet is an anticyclonic rotation (clockwise, $\mathrm{Ro}<0$) centered at the position of the vortex and extending azimuthally for almost 1/4 of the disk between about 160\arcdeg\, and 250\arcdeg. In this region, the radial and azimuthal  velocity of the gas varies between $\pm 0.5$ km s$^{-1}$ and  $\pm 0.3$ km s$^{-1}$, 
respectively. The velocity deviations from Keplerian motion are comparable to the local sound speed and, as discussed in \cite{li2001rossby}, might produce shocks capable of forming spiral density waves themselves. For the purpose of this paper, a key aspect is that the velocity perturbations associated with the vortex are sufficiently large to be detected with spatially resolved spectroscopic observations of molecular line emission from protoplanetary disks. The detectability of the kinematic signatures of a vortex, and, in general, of the planet-disk interaction, is discussed in the following sections.

\subsection{Obtaining Disk Temperature Distributions}

\begin{figure}[!t]
\centering
\includegraphics[width=0.5\textwidth]{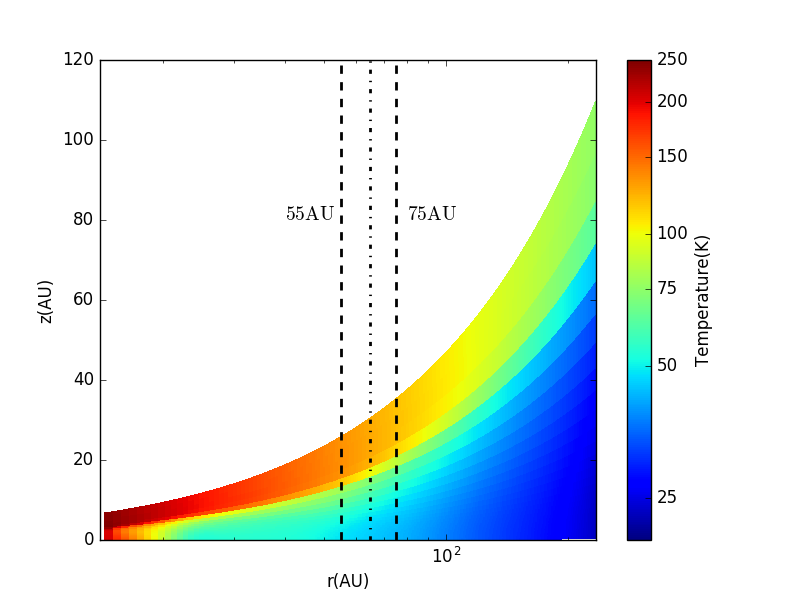}
\caption{Azimuthally averaged disk temperature obtained through 3D radiative transfer calculations. The inner radius and outer radius of vortex are marked by the dashed lines. The radial location(65 AU) of vortex center is marked by the dot-dashed line.}
\label{fig:figure3}
\end{figure}

We use the radiative transfer code RADMC-3D \citep{dullemond2012radmc} to calculate the temperature and emission of the disk perturbed by the planet following a procedure similar to that described in \cite{jin2016modeling}. In brief, we first convert the 2D gas surface density distribution generated by LA-COMPASS  into a 3D cylindrical density structure defined as $\rho(r,\phi,z) = \frac{\Sigma(r,\phi)}{\sqrt{2\pi}h(r)}e^\frac{-z^2}{2h(r)^2}$, where, as above, $h(r)= c_s(r)/\Omega(r)=h_0(r/r_0)^{5/4}$. Since the disk opacity at the wavelengths at which disk heating and cooling occur (i.e., optical and infrared) is controlled by (sub)micron grains that are coupled to the gas, it is appropriate to use the gas density for the calculation of the disk temperature.
To calculate the disk temperature we assume a disk opacity 
proper of interstellar dust made of astronomical silicates \citep{weingartner2001dust}, organic carbonates \citep{zubko1998size}, water ice, with fractional abundances as in  \cite{pollack1994composition}, and a grain size distribution $n(a)\propto a^{-3.5}$  between $0.01-1$ $\mu$m. We include instead grains as large as 1 mm to calculate the millimeter-wave dust emission. In both cases, the gas-to-dust ratio is set to 100.
We then perform a radiative transfer Monte Carlo simulation which leads to a Monte Carlo noise on the disk midplane temperature below 3\% as required to properly solve for the hydrostatic equilibrium \cite[see, e.g.][]{Isella18}. Since the disk density and temperature are mutually dependent, a few iterations are required to reach a stable solution. We find that the disk reaches stable hydrostatic and thermal equilibrium at a midplane temperature $T(r) \approx 73\textrm{K}(r/r_0)^{-1/2}$, or $h(r)=0.085 (r/r_0)^{5/4}$. The disk temperature is to the first approximation azimuthally constant (Figure~\ref{fig:figure3}), although the higher gas and dust density at the vortex causes a small azimuthal temperature variations (by a few degrees), whose effects might be difficult to detect. A more detailed discussion on the radial and azimuthal temperature perturbation caused by the disk-planet interaction can be
found in \cite{Isella18}.

We note that the 3D disk temperature obtained via RADMC-3D gives the typical two-layered structure as shown in Figure~\ref{fig:figure3} \citep[e.g.,][]{chiang1997spectral}. The radial 
temperature profile in the midplane, however, is lower
than the initial temperature profile used in the 2D
hydrodynamical simulations by about $20-30\%$. 
Furthermore, because the 3D disk temperature distributions
were obtained via post-processing radiative transfer calculations, these quantities are presumably not self-consistent
with the realistic dynamic radiation 
hydrodynamic calculations. 
The exact impact of these simplifications will need
to be understood in future studies.

\subsection{Synthetic images and mock-up ALMA observations}

Once the disk temperature is set, we calculate the synthetic disk emission in the $^{12}\textrm{CO}$ $J=3-2$ line emission at 345.796 GHz and in the surrounding dust continuum emission. To calculate the dust emission we assume that millimeter grains are settled toward the midplane with respect to the gas. Their vertical distribution is assumed to be Gaussian with a pressure scale height $h_{dust}(r)=0.1h(r)$ \citep{isella2016ringed,2016ApJ...816...25P}. The CO emission is calculated assuming LTE \citep[see, e.g.,][for a discussion on the applicability of LTE to protoplanetary disks]{2018ApJ...853..113W}  and a fractional abundance $\chi(^{12}\mathrm{CO})=1.4\times 10^{-4}$ 

relative to $\mathrm{H_2}$. The Doppler shift of the line emission is calculated using the gas velocity $\vec{v}_g$ provided by the hydrodynamic simulation discussed above (see Figure~\ref{fig:figure2}). The spectral profile of the emission line is calculated accounting only for thermal broadening, i.e., no turbulence is included in the model. This choice is motivated by the low values of turbulence measured in some protoplanetary disks \citep{2017ApJ...843..150F,2018ApJ...856..117F,2018arXiv180801768T}.
We generate synthetic images for disk inclinations of 25\arcdeg\, and 50\arcdeg\, in a velocity interval between $\pm$7 km s$^{-1}$. 

As the final step in our modeling procedure, we use the `simobserve' task in the Common Astronomy Software Applications package  \citep[CASA\footnote{\label{note1}https://casa.nrao.edu/}][]{mcmullin2007casa} to generate mock-up ALMA observations that account for the effect of finite angular resolution and sensitivity. Our goal is to investigate the detectability of the kinematic signatures of the vortex in the line emission and the factors that affect such signatures. Since the vortex has a spatial extent of approximately 30 AU, or 0.25\arcsec\, at a source distance of 120 pc, we adopt an array configuration that provides a FWHM beam size of about 0.1\arcsec\, with natural weighting (CASA antenna file number 17). Lower resolution images would not spatially resolve the vortex, while observing the CO emission at higher resolution with good sensitivity would require very long observations. 

The azimuthal position of the vortex is important because observations of the line emission measure the component of gas velocity along the line of sight. As a consequence, if the vortex is located along the apparent minor axis of the disk ($\theta=90$\arcdeg), the $v_\phi$ component of the gas velocity will mostly be perpendicular to the line of sight and will not generate any Doppler shift. In this model, the kinematic signature of the vortex would depend on the radial component of the gas velocity. Vice versa, when the vortex is close to $\theta=180$\arcdeg, its observable kinematic signature will depend on $v_\phi$, while the $v_r$ component along the line of sight would be close to zero. As such, we generate mock-up observations by rotating the synthetic models so that the vortex is either close to the projected minor axis of the disk ($\theta=128$\arcdeg) or to its major axis ($\theta=198$\arcdeg\ and $\theta=205$\arcdeg).

Beside the angular resolution of the observations and the location of the vortex, the detectability of the vortex kinematics would depend on the sensitivity and velocity resolution of the observations.  Here, we generate mock-up observations that provide velocity resolutions of 0.1 km s$^{-1}$ and 0.05 km s$^{-1}$, and sensitivities between 3.5 K and 5.5 K. 

In total, we generated a set of 5 mock-up observations whose observational parameters are listed in Table~\ref{tab:table1}. Each mock-up observation consists of an image of the dust continuum and a continuum subtracted channel maps  of the $^{12}$CO emission. 

\begin{figure*}[!t]
\centering
\includegraphics[width=\textwidth]{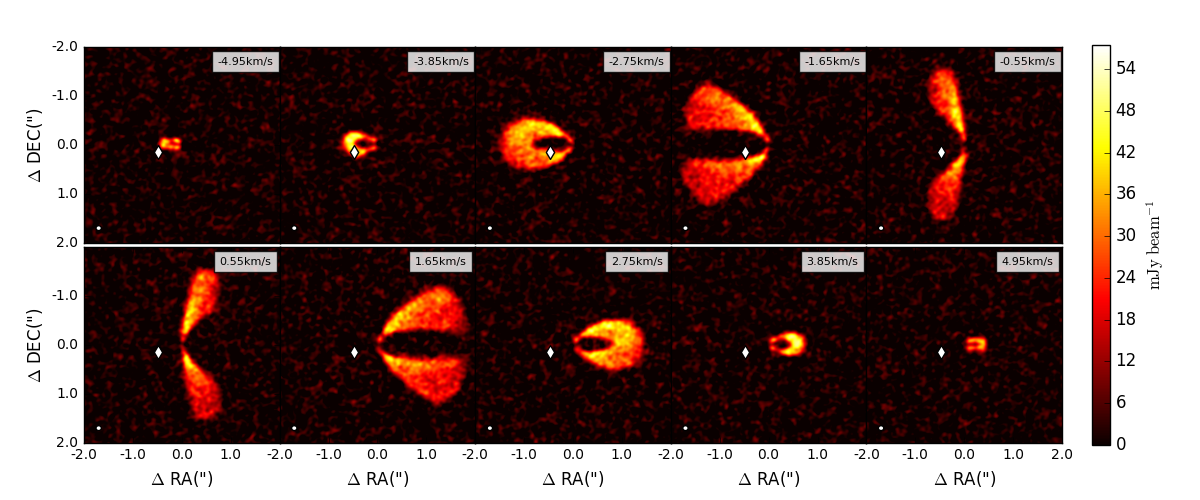}
\caption{Subset of the mock-up channel maps of the $^{12}$CO J=3-2 line emission for model A. The position of vortex is indicated with a diamond symbol, while the size of the synthesized beam is marked as a white ellipse in each panel. Our model covers a velocity interval from -7.0 km s$^{-1}$ to 7.0 km s$^{-1}$ at 0.1 km s$^{-1}$ velocity resolution, for a total of 140 channels. The x and y axises show the offsets relative to the central star, which is located at (0,0). }
\label{fig:figure4}
\end{figure*}

\begin{figure}[!t]
\centering
\includegraphics[width=0.5\textwidth]{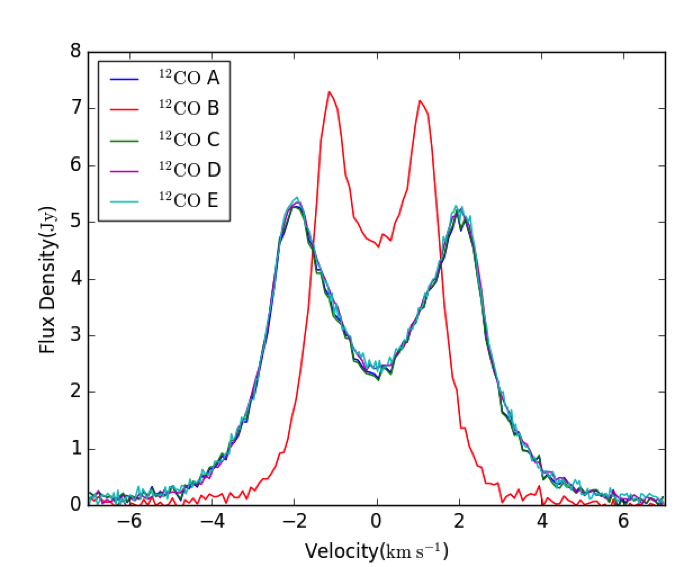}
\caption{Spectra of the $^{12}$CO J=3-2 line emission integrated across the entire disk for all five models.}
\label{fig:figure5}
\end{figure}

A subset of $^{12}$CO channel maps for model A are shown in Figure~\ref{fig:figure4}, while spatially integrated spectra 
for all models are presented in Figure~\ref{fig:figure5}. The peak of the line 
emission is optically thick ($\tau_{co}>500$) across most of the disk including the vortex, while the dust emission reaches an optical depth of about 5 only at the center of the vortex. The line spectra present the characteristic double peak shape caused by the disk rotation. The line emission of model B is narrower and more intense because of the smaller inclination. In general, the blue-shifted peaks have slightly higher intensities than the red-shifted peaks. As discussed in the next section, this is caused by the enhanced gas surface density in the vortex region. It is worth mentioning that we have investigated the observability of the vortex also in the $^{13}$CO J=3-2 line, but given the much lower signal-to-noise ratio of the simulated observations (or, the much longer telescope time required to achieve the same signal-to-noise ratio of $^{12}$CO observations), they do not provide any advantage in detecting the kinematic signature of the vortex, and therefore are not discussed here. 

\begin{figure*}[htp]
\includegraphics[width=\textwidth]{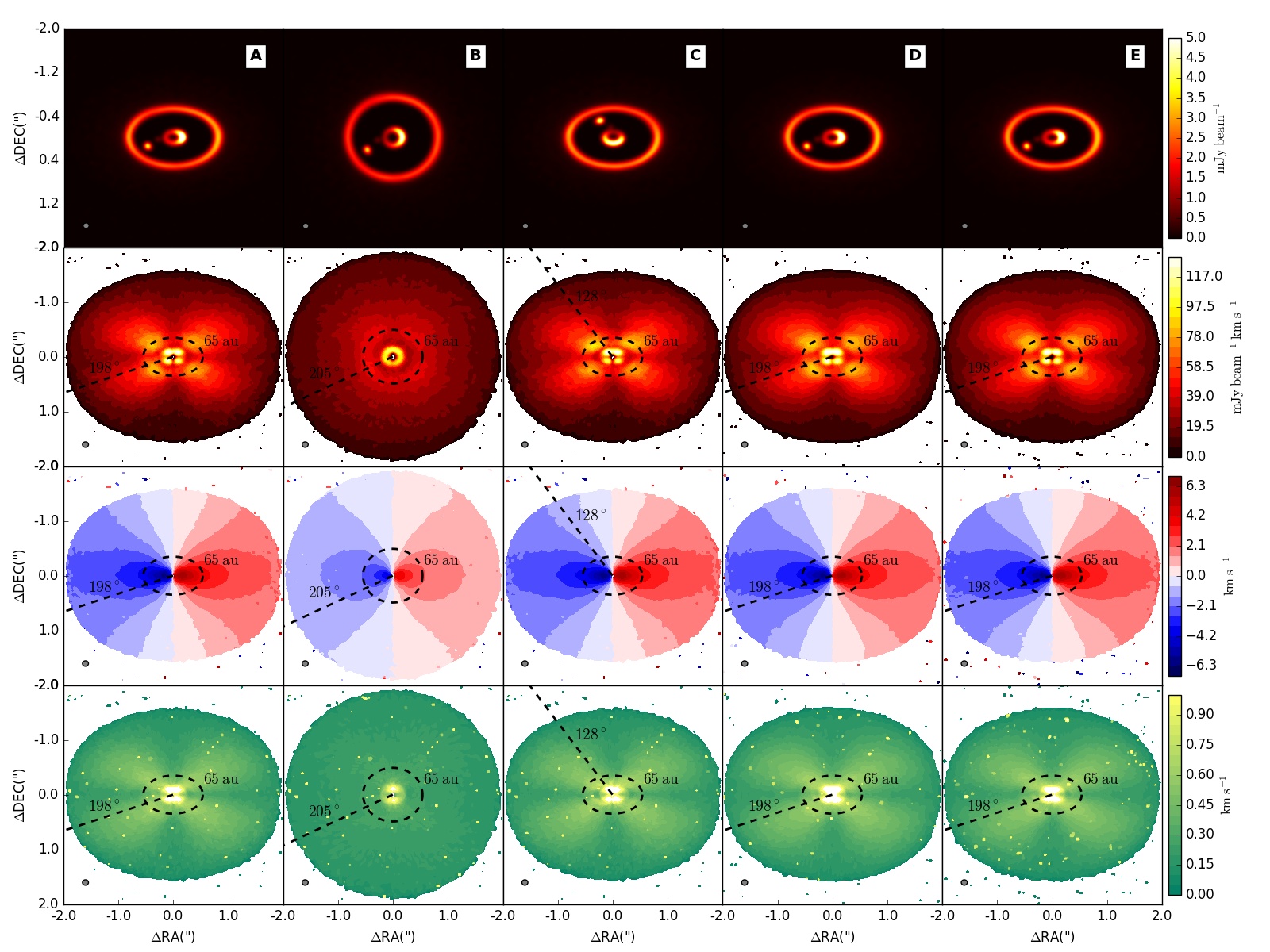}
\caption{The dust continuum (345.796 GHz) and $^{12}$CO J=3-2 line emission moment maps for model A to E
(columns). From top to bottom, each row presents the dust continuum, zeroth moment, first moment and second moment map of gas intensity of each model, respectively. The size of the synthesized beam is shown as a gray ellipse at the bottom left corner of each panel. Where the dashed ellipse and the dashed line intercept marks the center of vortex in each panel.}
\label{fig:figure6}
\end{figure*}

In addition to channel maps and spatially integrated spectra, we produced images of the zeroth, first, and second moment of the line emission, corresponding to the spectrally integrated line intensity, intensity weighted velocity centroid, and intensity weighted line width, respectively. These maps are shown in Figure~\ref{fig:figure6} together with the image of the dust continuum emission. 

The center of the vortex in the moment maps is marked as the intersection between the black dashed line and the black dashed ellipse. The zeroth and second moment maps of $^{12}$CO look like butterfly with cross-shape feature, because the inclination of disk lets the channels with larger velocity gradient along the line of sight have better relative velocity resolutions \citep{adam1990line}. 

In the following section we discuss the observability of the vortex in each of these models.

\begin{table*}\scriptsize
	\centering
	\caption{Model parameters of the synthetic ALMA observations.}
	\label{tab:table1}
	\begin{tabular}{lcccc}
		\hline
		Model & Inclination & Position Angle of Center of Vortex  & Integration Time & Velocity Resolution \\
		\hline
		A(standard run) 		& $50^\circ$ & 198\arcdeg   & 2 hr & 0.1 km/s  \\       
        B		 				& $25^\circ$ & 205\arcdeg   & 2 hr & 0.1 km/s  \\       
        C 						& $50^\circ$ & 128\arcdeg   & 2 hr & 0.1 km/s  \\     
        D		 				& $50^\circ$ & 198\arcdeg   & 4 hr & 0.1 km/s  \\
        E		 				& $50^\circ$ & 198\arcdeg   & 4 hr & 0.05 km/s \\

		\hline
	\end{tabular}
\end{table*}

\begin{figure*}[htp]
\includegraphics[width=\textwidth]{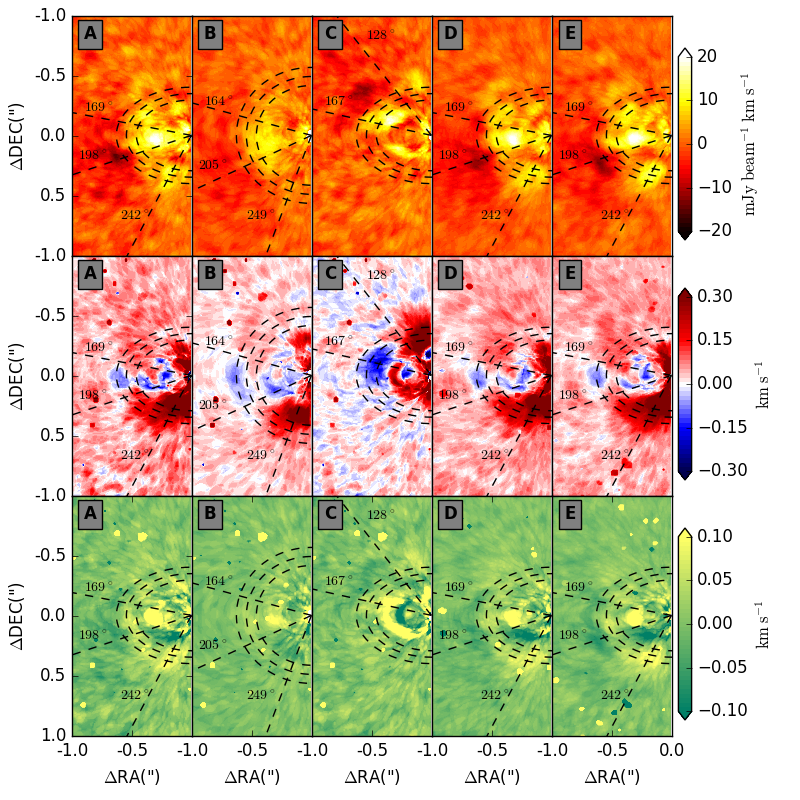}\centering
\caption{From top to bottom, left to right: Difference between the blue shifted side and the red shifted side of zeroth, first and second moment of $^{12}$CO J=3-2 line emission for each model. The vortex regions are marked as the zones between the black dashed ellipses and the black dashed lines. The dashed ellipses from inside to outside indicate radii of 55 au, 65 au, and 75 au same as Figure~\ref{fig:figure1}. The dashed lines mark the azimuthal position of vortex in the moment maps. We can see that there are blue and red deviations compared with the background between 55 au to 75 au in the middle row. These features are caused by the anticyclonic velocity field of a vortex. }
\label{fig:figure7}
\end{figure*}

\begin{figure*}[htp]
\includegraphics[width=\textwidth]{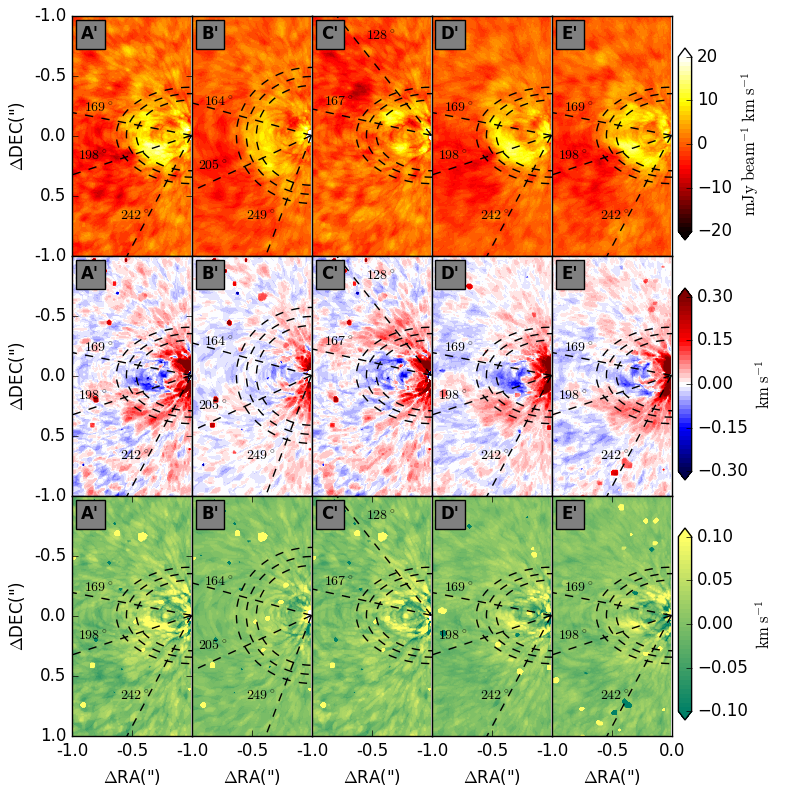}\centering
\caption{Same as the Figure~\ref{fig:figure7}. The surface density and temperature profiles of A' to E' are the same as A to E in Figure~\ref{fig:figure7}, but the velocity field input of RADMC-3D is now purely Keplerian velocity field. Compared with Figure~\ref{fig:figure7}, the blue and red deviations between 55 au to 75 au are much weaker in the middle row. Differences between
the two figures indicate that the anticyclonic velocity field of a vortex can be observationally detected. }
\label{fig:figure8}
\end{figure*}

\section{Kinematic signature of the vortex in ALMA mock-up observations}

At first glance, Figures 4--6 do not show any obvious signature of the vortex in the molecular 
line emission, despite the fact that it significantly perturbs the gas velocity. The first result of our 
investigation is therefore that vortices lead to rather small perturbations in the line emission and 
can hardly be identified by visually inspecting channel maps. 

Here we search for the faint vortex signature based on the following idea. In the model of an unperturbed Keplerian disk, 
channel maps with the same absolute value of the radial velocity are symmetric with respect 
to the apparent minor axis of the disk (which corresponds to the projection of the angular moment 
vector on the plane of the sky). This means that, besides the opposite sign of the radial velocity, 
the blue and red shifted sides of the disk should be similar. We can therefore search for kinematic 
perturbations from Keplerian rotation by folding the line emission along the disk minor axis and 
subtracting, pixel by pixel, the red shifted emission from the blue shifted line intensity. 
Figures~\ref{fig:figure7} shows this procedure applied to the moment maps of 
the ALMA mock up observations. Below we discuss the signature of the vortex in each of the five 
models listed in Table 1.

\subsection{Model A: Standard Run}
In this models, the disk has an inclination of $50^\circ$ similar to IRS~48 \citep{geers2007spatial} and the center of 
vortex is located at a position angle of 198\arcdeg. The observations have a velocity resolution of 0.1 km s$^{-1}$ 
and achieve a sensitivity of 5.2 K. In the zeroth moment map of $^{12}$CO, the intensity of 
gas emission in center of vortex is lower than in the nearby region. This is in conflict with the 
fact that the surface density of gas peaks at the center of the vortex (see Figure~\ref{fig:figure1}). 
This phenomenon was also reported by \cite{boehler2017close} in the HD 142527 CO gas line observations and is 
caused by the fact that the optically thick line emission arising from the disk surface absorbs the continuum 
emission originating from deeper layers of the disk. Continuum subtraction then leads to artificial removal 
of line emission \citep{boehler2018complex}. When both dust and gas emission are optically thick (as it happens 
at the center of the vortex), this situation become more severe and continuum subtraction might result in the 
removal of most of the line emission. In our model, however, the intensity of the $^{12}$CO emission only 
drops slightly due to continuum subtraction and the detectability of the vortex itself is not affected. We have tested this by generating images of the non-continuum subtracted line emission.
But since the results are very similar, we are not 
showing them here.

The difference between the blue shifted side and red shifted side of zeroth moment map of the $^{12}$CO $J=3-2$ line 
for model A is shown in the top row of Figure ~\ref{fig:figure7}. The region enclosed by the dashed circles and the dashed lines with position angle marks the region of vortex. Reading from the top panel of model A, the difference is negative outside the radius of the vortex and positive inside. This is due 
to the fact that the gas and dust density enhancement within the vortex changes how the disk is illuminated by the 
star, and consequently its temperature. In practice, the inner side of the vortex is more directly illuminated by 
the star and it is slightly hotter than the unperturbed disk. Vice versa, the outer part of the vortex is illuminated 
at a more slanted angle and it is slightly colder. 

The kinematic signature of the vortex can be seen by looking at the difference between the blue and red 
shifted sides of the first and second intensity moment. The vortex in the protoplanetary disk is an 
anticyclone, so it rotates clockwise compared with the anticlockwise rotation of the whole disk. 
This effect results in a detectable gradient in the velocity of the gas along the 
line of sight across the radial and azimuthal extent of the vortex. We can see that the blue shifted side of disk has a larger value of line of sight velocity than the red shifted side outside 75 au in all models,  as shown by the second row of Figure~\ref{fig:figure7} with the overall reddish background over the whole domain. (The implications of this effect will be 
explored in future studies.) On top of the reddish velocity difference background, however, there is a blue deviation($\sim 0.1\mathrm{km\;s^{-1}}$) between 55 au to 75 au, $\mathrm{169^\circ}$ to $\mathrm{198^\circ}$. This blue deviation is caused by the anticyclone vortex. The similar feature was also seen in the first moment maps of model B, D and E. In particular, gas located outward 
of the vortex center rotates slower than Keplerian and its velocity along the line of sight is about -0.15 km/s 
lower than the velocity at the center of the vortex. Vice versa, gas located inside the radius of the vortex 
center rotates faster than Keplerian, and its line of sight velocity is about 0.15 km/s faster than that 
of the center of the vortex. As a result, the velocity gradient along the vortex is about 0.3 km/s, i.e., about 
three times bigger than the velocity resolution of the observations. Note that such a velocity gradient 
can manifest itself both in the radial and azimuthal directions. 

The vortex also has a signature in the line width as described by the second moment of the line intensity, as shown in the bottom panel of Figure ~\ref{fig:figure7}. 
In this model, regions close to the center of the vortex have higher velocity dispersions which is caused 
by the fact that the observations do not spatially resolve the innermost part of the vortex and therefore 
the emission from gas regions moving at different velocities are all observed within one resolution element.  
At the resolution of 0.1$\arcsec$ this effect is however rather minor (the velocity dispersion increases 
by only 0.05 km/s). 

In order to further illustrate how the velocity fields affect the moment maps, we run another 5 models, which are marked as A' to E'. The surface density and temperature profile are kept the same as model A to E and we just changed the velocity input as purel Keplerian velocity field when we generate the synthetic images using RADMC-3D. Detailed comparisons between Figure~\ref{fig:figure7} and Figure~\ref{fig:figure8} show that: 
1) Even though some generic features such as the blue and red regions seen in the first moment maps are still visible in Figure~\ref{fig:figure8}, they are quite a bit weaker and not 
as systematic. 2) Features in the second momentum maps of
Figure~\ref{fig:figure8} become too weak to discern. 
These comparisons indicate that velocity fields containing a 
vortex made by a planet can indeed be ``visibly'' different from
a purely Keplerian velocity field, lending further support
of detecting such features in actual observations.

In conclusion, the combination of the perturbation signatures observed in the moment zeroth, first, and second, 
can pinpoint the position of the vortex and yield the magnitude of its rotational velocity.

\subsection{Model B: Effect of Disk Inclination}
In the model B, we lower the disk inclination from $50^\circ$ to $25^\circ$. Because of the lower disk 
inclination, the disk is more circular and the gas velocity along the line of sight is smaller. 
As a result, the  signature of the vortex in the gas moment maps becomes less significant than in model A. It is obvious that the blue and red deviations are less significant than the model A ($\sim 0.75 \mathrm{km\;s^{-1}}$). 
Lowering further the disk inclination would lead to a very weak and undetectable kinematic signature of the 
vortex. In the limiting model of a face-on disk, the component of the gas rotational velocity along the line 
of sight vanishes and the vortex becomes invisible. We compare the deviation of moment maps in B and B', we find that it is hard to see the the vortex features in B' because of the low inclination and the Keplerian velocity field in B'. This confirms that the low inclination is not favorable to find vortex features in the 2D hydrodyanmic models. In reality, vortices might cause a vertical circulation of the 
gas and perhaps they might be observed even at face-on inclination. However exploring this scenario would 
require 3D hydrodynamic models. In the opposite situation when the disk inclination is larger than $60^\circ$, the molecular line emission 
arising from the inner disk regions might be blocked by the outermost flaring disk regions. This would also hamper 
the investigation of the vortex kinematics. Our analysis shows therefore that the best inclination range to detect kinematics signatures of the vortex spans between 30$^\circ$ and 60$^\circ$.

\subsection{Model C: Effect of the Azimuthal Position of Vortex}
Because observations measure the component of the gas velocity along the line of sight, the azimuthal position 
of the vortex with respect to the apparent minor axis of the disk has an effect on its observability. In particular, velocity 
perturbations with respect to the Keplerian motion of vortices located close to the disk major axis (PA$\sim0^\circ$,$180^\circ$) are proportional to $V_\phi \sin(i)$, while velocities perturbations of vortices located close to the disk minor axis (PA$\sim90^\circ$,$270^\circ$) depend on $V_r \sin(i)$, where $V_\phi$ and $V_r$ are the tangential and radial velocities, respectively (Figure~\ref{fig:figure2}). In model C, we rotate the viewing angle so that the center of the vortex is at a position angle of 128$^\circ$. Despite the fact that part of the vortex now extends in the red-shifted part of the disk, its kinematic signature remains clearly visible in the first moment difference. Actually, the amplitude of the velocity difference is larger compared to model A and varies between $\pm$0.25 km/s. 

\subsection{Model D: Effect of Thermal Noise}
The thermal noise level of the observation is proportional to $1/\sqrt{t}$, where $t$ is the 
observation time. To investigate the effect of thermal noise on the observations, in model D we increase the integration time 
on source from 2 hr to 4 hr. The rms noise per channel consequently drops from 2.9 $\mathrm{mJy \; beam^{-1}}$ in model A to about 2.1 $\mathrm{mJy \; beam^{-1}}$. Because the noise level has dropped by about 30\%, the dust continuum and gas moment maps in model D achieve higher signal-to-noise and appear smoother compared to model A. However, the difference between blue shifted and red shifted part of the first and second moments changes only slightly. The comparison between D and D' is similar to the cases between A and A', but just less noisy.

\subsection{Model E: Effect of Velocity Resolution}
In model E, we increase the velocity resolution from 0.1 km/s to 0.05 km/s. Since the rms noise is proportional to $1/\sqrt{\Delta v}$, we extend the total observation time to 4 hours in order to compare with our standard run.
The dust continuum and moment maps are shown in the fifth column of Figure~\ref{fig:figure6}, while differences between the blue and red shifted sides of the disk are presented in the fifth column of Figure~\ref{fig:figure7}. The comparison of E and E' in Figure~\ref{fig:figure7} and Figure~\ref{fig:figure8} are similar with A and A'. Overall, the kinematics signature of the vortex  in model E is very similar to that observed in model A and D. The higher velocity resolution results in a smoother map of the second moment but the improvement is marginal and perhaps not worth the longer integration time.

\section{Discussion and Conclusions}
\label{conclusion}

ALMA observations have revealed azimuthal asymmetries in several protoplanetary disks that are 
thought to trace large scale vortices produced by the Rossby Wave Instability excited by the 
planet-disk interactions. If this interpretation
is correct, such vortices should perturb the motion of the gas, and consequently, the spectral profile
of the gas emission from the affected regions. To investigate whether these perturbations could
be detected, we have performed high-resolution 2D hydrodynamic and 3D radiative transfer simulations 
of a protoplanetary disk characterized by the presence of an anticyclonic vortex
generated by the gravitational interaction between the disk and a 10 Jupiter mass planet located 
at 35 au from the central star. The stellar and planet properties were chosen to roughly match
those of the IRS~48 system which is characterized by the most extreme azimuthal asymmetry discovered 
to date. The vortex traps the dust as previously discussed by \cite{fu2014long} and produces a compact 
azimuthal asymmetry characterized by an azimuthal density variation larger than three orders of
magnitude. As a reference, the azimuthal dust density variation measured in IRS~48 is larger than a
factor of 100. By comparison,  the gas density varies azimuthally only by a about a factor of two (Figure~\ref{fig:figure1}). 
In addition to the dust trapping, the vortex induces variations in the gas velocity of the order 
of the gas sound speed (Figure~\ref{fig:figure2}) which can be probed be observing the molecular line 
emission. 

We used CASA to generate images of dust continuum and $^{12}$CO J=3-2 emission around the frequency 
of 345 GHz, that could be observed in ALMA band 7. We chose this band because it provides the 
best sensitivity to thermal gas and dust emission. We explored the observability of the vortex
kinematic signature by varying observational parameters such as the disk inclination and position angle,
the integration time, and the velocity resolution.  

Overall, we found that despite the rather large perturbations in the gas velocity 
(Figure~\ref{fig:figure2}), the signature of the vortex on the $^{12}$CO line emission 
is weak, and it can be easily overlooked by visually inspecting the channel maps or the 
intensity moments. This is due to the fact that (i) the azimuthal variation in the gas 
density at the radius of the vortex is much weaker than the azimuthal variation in the dust 
density, and (ii) the $^{12}$CO line is mostly optically thick and therefore only weakly depends
on the gas density. Nevertheless, we show that the velocity perturbations due to the vortex 
can be recovered by comparing the blue and red-shifted sides of the disk. 
We find, in particular, that the rotation of the vortex is clearly visible by folding 
the first moment map along the disk apparent minor axis and subtracting the blue-shifted 
emission from the red-shifted emission, or vice versa. This is a simple approach that 
does not require performing any detailed and time consuming forward modeling. 

In order to be more 
concrete, we have used the disk parameters of IRS 48 system as a case study to investigate the 
detection feasibility by ALMA. By exploring different observational conditions, we established the following recipe to maximize 
the probability to directly detect vortices in protoplanetary disks using ALMA observations. 

\begin{enumerate}
\item To detect the kinematic signature of a vortex, it is necessary to spatially resolve it 
across a few resolution elements. Although a careful analysis of the 
spatially integrated line profile might also reveal velocity perturbations, our
simulations show that the effect on spatially integrated spectra is generally too weak 
to be unambiguously attributed to a vortex. Given the angular size of the dust crescents 
observed so far, we find that the observations should achieve an angular resolution of about
0.1$\arcsec$, corresponding to spatial scales of about 10-20 au at the distance of nearby star
forming regions. 

\item To measure the velocity gradient across the vortex, the observations should detect the line emission 
with a peak signal-to-noise ratio larger than 15. Because the temperature of protoplanetary 
disks at a few tens of au from the central star ranges between 20-100 K, the observations should 
achieve a brightness temperature sensitivity of 1-7 K if the line under consideration is optically 
thick and it fills the beam entirely (i.e., there is no beam dilution). 
Achieving such a sensitivity with a beam size of 0.1$\arcsec$ and a velocity resolution of 0.1 km/s would 
require between about 1 hr (to reach an rms of 7 K) and 48 hr (to reach an rms of 5 K) of integration time 
on source with ALMA. Observing disks around hot stars is therefore highly preferable. 
Furthermore, increasing the angular resolution to 0.08$\arcsec$ would imply incrementing the observing 
time by a factor of 2.5, therefore making such an observational program hardly feasible.  
Observations of more optically thin lines, such as, for example, $^{13}$CO J$=3-2$ could potentially
better trace the perturbations in the gas density caused by the vortex. However, we find that achieving
the required s/n ratio on this line would require more than 15 hr of integration time on source.  

\item To measure the velocity gradient across the vortex, the observations should have a velocity 
resolution at least 3$\times$ higher than the amplitude of the velocity perturbations induced by the vortex. 
We find that the velocity variations associated with the vortex can be comparable to the sound speed in the gas, 
which, for typical temperatures values, varies between 0.3
-0.6 km/s. Our simulations show that a velocity resolution of 0.1 km/s provides the best compromise
between the need to spatially resolve the vortex kinematics and those of achieving a sufficiently high 
S/N ratio. 

\item Finally, we find that the disk inclination plays an important role in the capability to investigate the vortex kinematics. A low disk inclination ($i < 30\arcdeg$) will cause a smaller projection of the gas velocity along the line of sight and weaker vortex signatures. A high disk inclination $(i> 60\arcdeg)$ 
might cause the emission from the vortex to be absorbed by the outermost disk regions. The best target candidates should therefore be those disks characterized by moderate inclinations. In the same time, we find that the azimuthal position of vortex in the moment maps is related to the detection the asymmetric feature in the gas line moment and the effect is still unclear. This may be figured out in our future work.
\end{enumerate}

One future study is to investigate the signatures of vortices in a background disk flow which is eccentric. Such configurations can be produced if the planet is quite massive or the disk contains a binary system.

\acknowledgments
We thank the referee for the detailed comments which improve the presentation of this paper significantly.
Part of this work was performed at the Aspen Center for Physics, which is supported by National Science Foundation grant PHY-1607611.
This work is supported by National Natural Science Foundation of China (grant Nos.11773081, and 11661161013), CAS Interdisciplinary Innovation Team, the Foundation of Minor Planets of Purple Mountain Observatory. The financial support provided by the UCAS (UCAS[2015]37]) Joint PhD Training Program and LANL/CSES are acknowledged. A.I. acknowledges
support from the National Science Foundation through the grant No. AST-1715719 and support from the National Aeronautics and Space Administration through the grant No. NNX15AB06G.
H.L. and S.L. acknowledge the support from the National Aeronautics and Space Administration through the grant 
No. 80HQTR18T0061 and Center for Space and Earth Science at LANL. 

\software{LA-COMPASS \citep{li2005potential,li2008type}, RADMC-3D v0.41 \citep{dullemond2012radmc}, CASA ALMA pipeline 5.0.0 \citep{mcmullin2007casa}}

\bibliographystyle{aasjournal}
\bibliography{references}


\end{document}